\documentclass[runningheads]{llncs}

\usepackage[T1]{fontenc}   
\usepackage{graphicx}      
\usepackage{float}         
\usepackage{pdflscape}  
\usepackage{adjustbox}  

\usepackage[colorlinks=true, allcolors=blue]{hyperref}  
\usepackage{color}

\usepackage{url}           
\usepackage{array}         
\usepackage{comment}
\usepackage{amsmath}
\usepackage{moresize}

\makeatletter
\renewcommand\section{\@startsection{section}{1}{\z@}%
                       {-8\p@ \@plus -4\p@ \@minus -4\p@}%
                       {6\p@ \@plus 4\p@ \@minus 4\p@}%
                       {\normalfont\large\bfseries\boldmath
                        \rightskip=\z@ \@plus 8em\pretolerance=10000 }}
\renewcommand\subsection{\@startsection{subsection}{2}{\z@}%
                       {-8\p@ \@plus -4\p@ \@minus -4\p@}%
                       {6\p@ \@plus 4\p@ \@minus 4\p@}%
                       {\normalfont\normalsize\bfseries\boldmath
                        \rightskip=\z@ \@plus 8em\pretolerance=10000 }}
\renewcommand\subsubsection{\@startsection{subsubsection}{3}{\z@}%
                       {-4\p@ \@plus -4\p@ \@minus -4\p@}%
                       {-1.5em \@plus -0.22em \@minus -0.1em}%
                       {\normalfont\normalsize\bfseries\boldmath}}
\makeatother

\usepackage[skip=2.5pt, indent=15pt]{parskip}

\usepackage{caption}
\usepackage{listingsVDM}
\usepackage{overturelanguagedef}

\begin{document}

\title{A System-of-Systems Case Study for the Verification of Composed Digital Twins}
\titlerunning{SoS Case Study for Verification of Composed DTs}

\author{Mennatullah T. Khedr\inst{1}\orcidID{0009-0007-0096-3375} \and Mengwei Xu\inst{1}\orcidID{0000-0003-4978-3061} \and
John S. Fitzgerald\inst{1}\orcidID{0000-0001-7041-1807}
\and Peter Gorm Larsen\inst{2}\orcidID{0000-0002-4589-1500}}

\authorrunning{M T Khedr, M Xu, J S Fitzgerald and P G Larsen}

\institute{School of Computing, Newcastle University, UK \\
\email{m.t.y.a.khedr2@newcastle.ac.uk, mengwei.xu@newcastle.ac.uk, john.fitzgerald@newcastle.ac.uk}
\and
Dept. of Electrical and Computer Engineering, Aarhus University, DK \\ \email{pgl@ece.au.dk}
}

\maketitle

\begin{abstract}
Current approaches to engineering dependable Digital Twins (DTs) of Cyber-Physical Systems lack practical guidance on how qualities such as relevance, verifiability, substitutability and fidelity may be formalised and verified. This need is amplified in Systems-of-Systems (SoS), where reliance is placed on the composition of DTs. 
The goal of this study is to identify foundational challenges that a framework for DT validation and verification should address in an SoS setting, and to investigate the formalisation of individual DT artefacts as a step towards compositional reasoning. 
We present a case study based on a DT-enabled greenhouse SoS modelled in VDM-RT (Vienna Development Method, Real-Time), including executable formal models, a property-based account of DT qualities, and an analysis of the obstacles arising when attempting to compose these artefacts at the SoS level. 
We consider how DT qualities may be operationalised as sets of verifiable properties. Formal modelling and analysis techniques and tools supporting their verification are identified. 
The study reveals how interpretations of DT qualities must adapt to different architectural roles and how local quality guarantees form assumptions for compositional reasoning. Full formalisation and verification of SoS composition is identified as future work.

\keywords{Digital Twins \and Cyber-Physical Systems \and Systems of Systems \and VDM-RT \and Formal Verification \and Digital Twin Qualities.}
\end{abstract}

\section{Introduction}
\label{sec:introduction}

Digital Twins (DTs) are synchronised digital representations of cyber-physical systems offering monitoring, prediction, and decision support~\cite{ISO_DTConcepts30173_2023,lv2022digital}. They can enable reasoning beyond direct sensing, supporting predictive maintenance, optimisation, and what-if analysis~\cite{tao2019digital,fuller2020digital,fitzgerald2024engineeringDTs}. Digital Shadows (DSs) differ from DTs in that they have a one-way, real-time data flow from their physical counterpart, without autonomously influencing the physical system. Where DSs and DTs influence or even perform critical functions, particularly in cyber-physical systems (CPSs), dependability becomes paramount.
There is therefore a need to ensure the quality of DTs to manage the risk of unsafe or costly impacts on the physical world~\cite{waters2025tevv,nationalacademies2024foundational}.

The risk of DT failure is amplified in System-of-Systems (SoS) settings. As DT-enabled systems scale, individual DTs interact with other DTs, shared digital services, and coordination mechanisms, forming interconnected architectures with SoS characteristics~\cite{olsson2023systems,borth2019DT_SoS_4Ch_4Arch}. Because DTs are developed under heterogeneous assumptions and different abstraction levels in SoS, 
faults in one DT can propagate to others that depend on it, making reasoning about the quality of composed DTs harder than for isolated DTs~\cite{incose2023handbook,JudithMichael2022IntegChall4DTSoS}.




The engineering of dependable DTs demands methods and tools for rigorous verification and validation, so that evidence can be provided to substantiate claims about DT qualities. 
However, recent surveys suggest that this aspect remains under-explored \cite{waters2025tevv,khedr2025composition,nationalacademies2024foundational,bitencourt2024verification}. Existing frameworks clarify terminology but are largely descriptive, and there remains a need for practical guidance on how DT qualities, such as relevance, verifiability, substitutability, and fidelity, should be formalised and verified~\cite{waters2025tevv,Oakes2023Qualities}. Consequently, such qualities are often assessed informally or at a single abstraction level, with limited treatment of how verification obligations change under SoS composition~\cite{Oakes2023Qualities,olsson2023systems}.


Key representative approaches to reasoning about DTs include contract-based prediction error detection~\cite{naeem2023contractbased}, TLA-based orchestration specification~\cite{huang2024formal}, and runtime monitoring for adaptive systems~\cite{hansen2024monitoring}. These demonstrate the feasibility of formal or semi-formal reasoning about DTs, but target isolated concerns, e.g., a single quality in~\cite{naeem2023contractbased} or coordination correctness in~\cite{huang2024formal}. 
Little work to date proposes how to formalise DT qualities, associates qualities with explicit verification obligations, or considers how those obligations change when DTs are composed in an SoS. At the same time, work on SoS-oriented DT development highlights interoperability and integration challenges~\cite{olsson2023systems,david2024interoperability}, yet offers limited guidance on how DT qualities should be evidenced under SoS composition. 

In response to the gaps discussed above, we propose to \emph{operationalise} DT qualities by interpreting them as sets of formally stated properties that can be verified or refuted with machine support. 
We investigate the approach through a case study of a greenhouse-inspired System of DT-enabled Plant Pots, comprising multiple plant-pot constituent systems, each enabled with a predictive DT, and supported by a shared Environment Digital Shadow, modelled in VDM-RT (Vienna Development Method, Real-Time). VDM-RT brings opportunities for both well-founded proof and extensive simulation and co-simulation. 

Our contributions are: (1) a formally executable DT-enabled greenhouse case study in VDM-RT supporting co-simulation; (2) an initial property-based approach to DT qualities with explicit verification evidence; (3) an investigation of how qualities manifest differently as verifiable properties across elements of DTs; (4) identification of verification challenges arising from DT composition in SoS settings.

\begin{figure}[h!]
    \centering
    \includegraphics[width=0.9\textwidth]{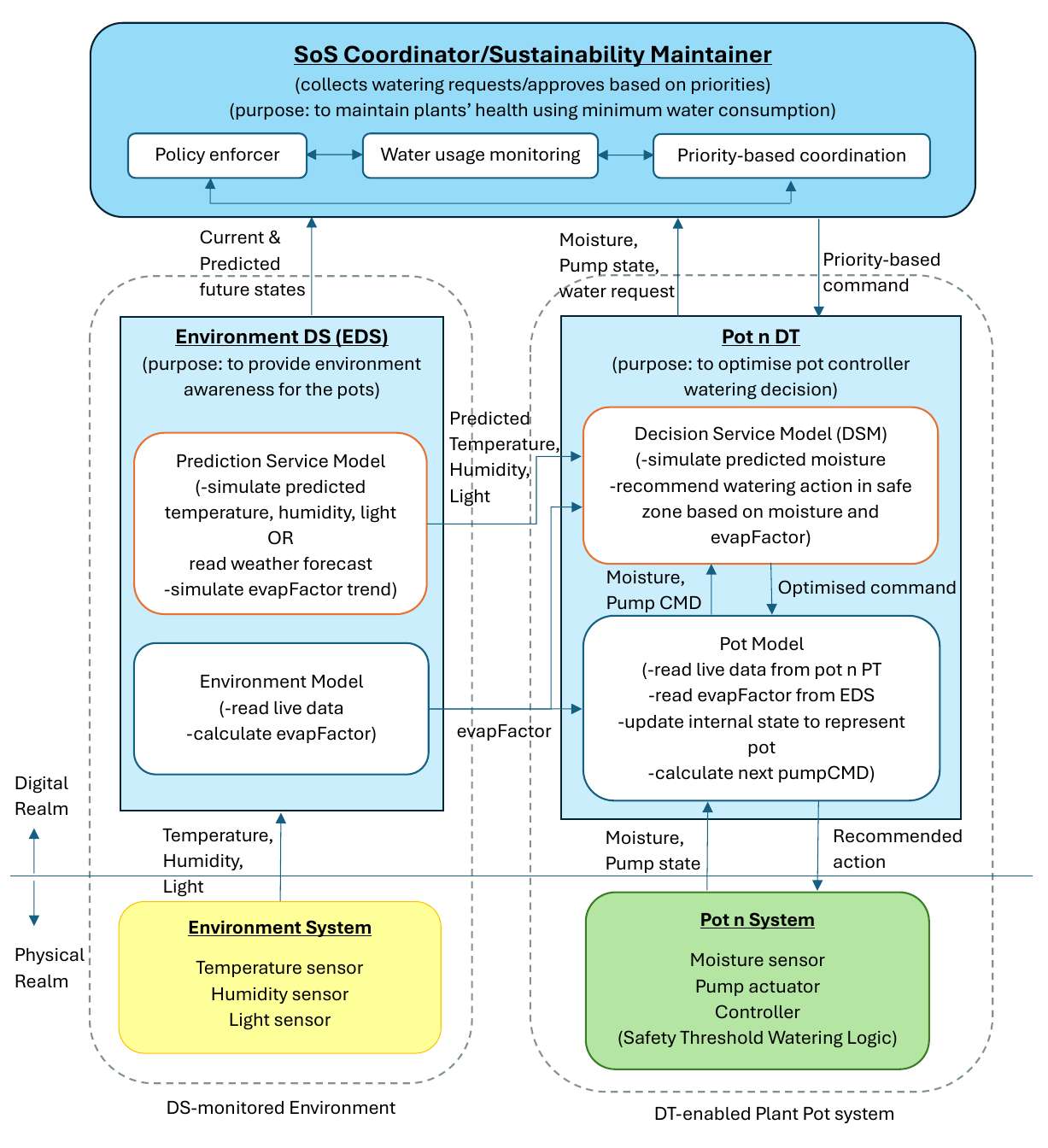}
    \caption{Architecture of the DT-enabled greenhouse SoS, illustrating physical systems, individual DTs, a shared Environment DS, and SoS-level coordination.}
    \label{fig:GreenhouseSoS_architecture}
\end{figure}

\textbf{Scope and limitations of this paper.} The contributions listed above focus on the formalisation of individual DT artefacts (the Pot Model, Decision Service Model, and Environment Digital Shadow) and on the qualities they instantiate. We do not present a full formalisation of the SoS‑level coordination layer shown in Figure~\ref{fig:GreenhouseSoS_architecture}, nor do we verify any composed properties across multiple DTs. Instead, Section~\ref{sec:compositionIssues} analyses the verification challenges that arise when moving from verified constituents to an SoS‑level composition - challenges such as fidelity error amplification, scope‑shifting relevance, and the evolution of substitutability into relational compatibility. These analyses are intended to motivate and inform future work on compositional verification, not to demonstrate compositional verification itself. Thus, this paper should be read as a foundational study of DT quality operationalisation and a scoping of the compositional verification problem, rather than as a complete solution for SoS‑level DT verification.

The remainder of the paper is structured as follows. Section~\ref{sec:case_study} introduces the case study. Section~\ref{sec:VDMformalModelling} describes VDM-RT modelling and co-simulation support. Section~\ref{sec:DT_qualities_properties} considers how DT qualities may be operationalised as verifiable properties. Section~\ref{sec:compositionIssues} examines compositional verification challenges. Section~\ref{sec:discusssion} discusses our key findings. Section~\ref{sec:validity} discusses threats to validity. Section~\ref{sec:conclusion} concludes and proposes future directions.

\section{Greenhouse System as a DT-enabled SoS}
\label{sec:case_study}

This section introduces the DT-enabled SoS for a greenhouse system that serves as a case study. The focus here is on the system structure and DT artefacts.

Figure~\ref{fig:GreenhouseSoS_architecture} presents the overall architecture of the DT-enabled greenhouse SoS. 
This case study is a minimal example of an SoS setting, which, despite its simplicity, is sufficient to allow the implementation and investigation of different SoS configurations and serves to demonstrate the difficulty of using DTs in such settings. Moreover, the architecture is explicitly structured to support hierarchical reasoning, the interdependence of individual plant pot systems and their DTs to SoS-level coordination.

\subsection{Physical Systems}

\subsubsection{Plant Pot System} Each plant pot system comprises a soil moisture sensor, a pump actuator, and a local controller that enforces safety-threshold logic. The controller accepts watering recommendations from its corresponding Pot DT only when the moisture is within a safe zone. Without DT support, this baseline strategy is purely reactive and unaware of future evaporation, shared resource constraints, or interactions with neighbouring pots.

\subsubsection{Environment System} The environment system consists of a controller connected to a temperature, humidity, and light sensor. Environmental data is continuously forwarded to an Environment Digital Shadow (DS), ensuring clear separation between shared context sensing and local control in the plant pots.

\subsection{Digital Artefacts (DTs, DS, SoS coordination)}

\subsubsection{Environment Digital Shadow}
The Environment DS (EDS), synchronised with the environment system, provides contextual information to the Pot DTs. It supports two modes of operation: a live mode that reads current sensor measurements and computes an evaporation factor, and a simulation mode that predicts environmental trends based on simplified models or external forecasts. This aligns with established notions of DSs as context-providing artefacts rather than decision-making entities~\cite{fuller2020digital}.

\subsubsection{Pot Digital Twin}
Each Pot DT provides environment-aware decision support to its corresponding plant pot system. It contains a Pot model that updates its state synchronously with real-world data and computes the expected threshold-based watering action, and a Decision Service model (DSM) that enables predictive reasoning about future moisture evolution, supported by environmental data, to generate refined watering recommendations for the physical pot controller. Importantly, the Pot DT does not directly actuate the system; safety enforcement remains the responsibility of the physical pot controller. This enables proactive and predictive optimisation while preserving safety guarantees.

\subsubsection{SoS-level Coordination DT}
At the SoS level, a composed coordination DT implements efficient priority-based watering strategies by aggregating water usage and enforcing global resource constraints across all pots through their Pot DTs. This layer enables reasoning about collective behaviour, such as peak water demand and fairness, that cannot be addressed by individual DTs alone.


To summarise, in this architecture, each physical plant pot controller enforces threshold-based safety locally. We then introduce DT functionality in two steps: (i) each pot is paired with a Pot DT, contextualised by a shared EDS; and (ii) a composed DT is created for SoS-level coordination. In this paper, we formalise and verify only step (i). We then discuss verification challenges for step (ii) in Section~\ref{sec:compositionIssues}, but leave its full formalisation for future work. Thus, the SoS‑level coordination layer serves as a target architecture and a source of research questions, not as a verified artefact in the present study.


\section{Formal Modelling with VDM-RT and FMI}
\label{sec:VDMformalModelling}

\subsection{Why VDM-RT for DT-enabled CPS?}

VDM-RT was selected as the primary formal modelling language due to its suitability for modelling DT-enabled CPSs that combine discrete control logic and data-centric state representations, and its large executable subset. VDM-RT supports explicit modelling of state variables, operations, invariants, and pre- and post-conditions, which are essential for expressing DT qualities as verifiable properties. The executable subset of VDM-RT enables early empirical validation via simulation and execution traces, while maintaining a rigorous formal foundation for subsequent verification~\cite{Verhoef&06b}. In addition, VDM-RT integrates naturally with FMI-based workflows, supporting the deployment of DT models as Functional Mock-up Units (FMUs) and their integration into co-simulation environments~\cite{Thule&18} for multi-artefact DTs and for composed DTs of SoS.

\subsection{VDM-RT Models Developed}

The current modelling effort focuses on the DT-enabled plant pot as the smallest unit of formalisation. VDM-RT models have been developed for three aspects. First, the Pot model, which is the behavioural abstraction of the physical pot (controller, sensor, and actuator), represents its moisture dynamics and safety-threshold control. Second, the DSM is a DT service that reasons about pot behaviour and produces an optimised, environment-informed recommended action to the physical pot controller. Third, the EDS models environmental dynamics and provides contextual information for relevant DTs.

The Pot class of the VDMRT Pot model consists of 183 lines of code (LOC) apart from the supplementary FMI-supporting files, such as the HardwareInterface, System, and World files, with a total of 244 LOC. The PotDSM.vdmrt class consists of 133 LOC, and the Environment.vdmrt class consists of 92 LOC. After running Proof Obligations Generation and checking them using QuickCheck, we found that the main functions and operations, considering the functionality of the models, were provable, while the operations that included reading from FMI ports were unchecked. This is because runtime-determined port values are unknown to the verifier, leading to uninterpreted terms and unchecked POs. For example, 56 POs were generated for the Pot model files, out of which 49 POs were for the Pot.vdmrt class and nearly half of them were unchecked due to FMI ports.getValue() calls.


\subsection{From VDM-RT to FMUs and Co-simulation}

The VDM-RT models are exported as FMUs in compliance with the FMI 2.0 standard. The export requirements impose constraints on the model structure, including explicit time handling, total operations, and the specified declaration of shared states via input and output ports\footnote{\url{https://into-cps-association.github.io/constituent-model-development/overture/fmi-support.html}}.

\begin{vdmtt}
class Pot
instance variables
  -- input ports
  public moistureSensor   : RealPort;
  public pumpState        : BoolPort;
  public evapFactor       : RealPort;
  public optimizedCommand : IntPort;  -- from PotDSM
  
  -- output ports
  public moisture_out     : RealPort; -- to PotDSM
  public pumpCMD          : BoolPort;
\end{vdmtt}

Initial co-simulation experiments demonstrate the executability and architectural soundness of the approach. Integration with RabbitMQ-based FMUs\footnote{\url{https://github.com/INTO-CPS-Association/fmu-rabbitmq?tab=readme-ov-file}} is ongoing, enabling future coupling with live data streams \cite{Frasheri&21b}. At this stage, co-simulation is primarily used to validate model behaviour and interaction patterns rather than for full operational deployment.

\section{DT Qualities as Structured Sets of Verifiable Properties}
\label{sec:DT_qualities_properties}

DT qualities, such as fidelity or relevance, are often described at a conceptual level without guidance on how they can be rigorously verified in practice. In our work, we consider how DT qualities can be operationalised as structured sets of rigorously defined 
and verifiable properties, starting from the interpretation of Oakes et al.~\cite{Oakes2023Qualities}. 
In this approach, we distinguish: 
\begin{enumerate}
    \item \textbf{Conceptual qualities:} high-level attributes of interest;
    \item \textbf{Formal properties:} precise logical statements that define those qualities;
    \item \textbf{Verification evidence:} the means by which a property may be justified;
    \item \textbf{Artefact instantiation:} embedding of properties within concrete DT components.
\end{enumerate}
\noindent The idea behind these distinctions is that the verification evidence contributes to a proof that the properties, and subsequently the qualities, are satisfied. We may regard properties as basic elements of a structured argument (as in a safety case) supporting the higher-level quality claim. Further, we aim to provide actionable guidance on how to embed these necessary properties using modelling constructs in the various DT components.


\subsection{From Abstract Qualities to Formal Properties}

For each high-level DT quality, we propose a set of formal properties to characterise the quality for the DT-enabled System (DTeS) of interest. 
We focus on four core qualities identified in~\cite{Oakes2023Qualities}, where they are initially described on models and their ``lifting'' to DTs is outlined:
\begin{itemize}
    \item \textbf{Relevance:} a model is indeed based on the system it purports to describe.
    \item \textbf{Verifiability:} the ability to determine whether properties of interest are satisfied in a given model.
    \item \textbf{Substitutability:} 
    safe replacement capability under defined assumptions.
    \item \textbf{Fidelity:} quantitative behavioural closeness to the referent.
\end{itemize}

\subsection{Formalisation and Evidence: From Properties to Justification}

The formally expressed properties of a DT are the basis for building an argument that claims its high-level qualities are satisfied. In our case study, such properties will typically be expressed using VDM-RT constructs. For example, state invariants have a role to play in a Relevance argument; operation contracts (pre/post- conditions) are used to establish Verifiability by constraining inputs, preserving invariants, and supporting termination guarantees; relational trace predicates support behavioural comparison for Substitutability; and quantitative error bounds provide value in capturing closeness for a case evidencing that a certain level of Fidelity holds. The evidential basis for such formal statements will come from a range of sources, such as static proof, contract verification, trace comparison, or empirical calibration. 



\subsection{Artefact-Specific Instantiation}
Our case study comprises three DT-related artefacts: the Pot Model (physical abstraction) and the DSM (optimisation logic), forming the Pot DT, and the EDS (context abstraction). All three artefacts may be evaluated against our qualities of interest. However, the interpretation of each quality depends on the architectural role of each artefact. For example, Tables~\ref{Potqualities},~\ref{DSMqualities}, and~\ref{EDSqualities} summarise properties relating to each of these artefacts. The property number (column 1) indicates a quality to which each property may be relevant (R for Relevance, V for Verifiability, etc.). We give an informal description of each property and the forms of evidence that may be relevant to establishing it.

\begin{landscape}
    \begin{table}[h]
    \centering
    \small
    \caption{Pot model – Property Classification}
    \begin{tabular}{@{}|
    >{\raggedright\arraybackslash}m{0.8cm}|
    >{\raggedright\arraybackslash}m{2.4cm}|
    >{\raggedright\arraybackslash}m{8.9cm}|
    >{\raggedright\arraybackslash}m{3.6cm}|
    >{\raggedright\arraybackslash}m{3cm}|
    @{}}
    \hline
    \textbf{No.} & \textbf{Property} & \textbf{Formal Expression} & \textbf{Description} & \textbf{Evidence} \\
    \hline
    P-R1 & Moisture Bounds & $0 \leq moisture \leq 100$ & Moisture within physical limits & Static proof (state invariant) \\ 
    \hline
    P-R2 & Valid Parameter Domains & $\begin{aligned}
         & 0 < \text{soilType\_coeff} \leq  1 \hspace{0.5cm} \land \hspace{0.5cm} \text{surfaceArea\_m2} > 0 \hspace{0.6cm} \land \\
         & \text{soilCapacity} > 0 \hspace{1.45cm} \land \hspace{0.5cm} 0 \leq \text{infRate} \leq 1 \hspace{1.3cm} \land \\
         & 0 \leq \text{minT} \leq 100 \hspace{1.5cm} \land \hspace{0.5cm} 0 < \text{maxT} \leq 100 \hspace{1.2cm} \land \\
         & \text{minT} < \text{maxT} \hspace{1.75cm} \land \hspace{0.5cm} \text{pumpFlow\_L\_min} \geq 0 \end{aligned}$ & 
         Meaningful physical parameters & Configuration validation \\ 
    \hline
    P-V1 & Invariant Preservation & \(moisture(t) \in [0,100]\) & Update respects physical constraints & Static proof \\
    \hline
    P-V2 & Moisture Update Totality & SimulateMoisture() always terminates & Update always produces next state & Structural analysis \\ 
    \hline
    P-V3 & Control Logic Termination & ControlLogic() always terminates & Pot control loop is total & Structural analysis \\ 
    \hline
    P-S1 & Behavioural Agreement & \(sign(\Delta moisture_{PhysicalPot}) = sign(\Delta moisture_{PotModel})\) & Pot model prediction agrees with physical Pot & Relational trace comparison \\
    \hline
    P-S2 & Threshold Consistency & \((moisture < minT)_{PhysicalPot} \Leftrightarrow (moisture < minT)_{PotModel}\) & Pot model aligns with physical pot on watering necessity & Relational trace comparison \\ 
    \hline
    P-S3 & Command Agreement & \(pumpState_{PhysicalPot} = pumpCMD_{PotModel}\) & Pot model produces same command as physical pot & Relational trace comparison \\
    \hline
    P-F1 & Moisture Prediction Error & \(|moisture_{PotModel} - moisture_{PhysicalPot}| \leq \varepsilon\) & Quantifies closeness to physical pot & Empirical calibration + trace comparison \\ 
    \hline
    P-F2 & Environmental Sensitivity \& Drying Consistency & $\begin{aligned} 
        & evapFactor(t) \geq evapFactor(t-1) \hspace{0.5cm} \land \\ 
        & pumpCMD=false \Rightarrow moisture(t) \leq moisture(t-1) \end{aligned}$ & 
        Pot model responds to environment changes and dries correctly when not watered & Empirical calibration + trace comparison \\ 
    \hline
    P-F3 & Watering Effect Monotonicity & \(pumpCMD=true \Rightarrow moisture(t) \geq moisture(t-1)\) & Watering increases moisture & Simulation-based validation \\ 
    \hline
  \end{tabular}
    \label{Potqualities}
\end{table}
\end{landscape}

\begin{landscape}
    \begin{table}[h]
    \centering
    \small
    \caption{Decision Service Model (DSM) – Property Classification}
    \begin{tabular}{@{}|
    >{\raggedright\arraybackslash}m{1.3cm}|
    >{\raggedright\arraybackslash}m{2.5cm}|
    >{\raggedright\arraybackslash}m{7.6cm}|
    >{\raggedright\arraybackslash}m{4.1cm}|
    >{\raggedright\arraybackslash}m{3cm}|
    @{}}
    \hline
    \textbf{No.} & \textbf{Property} & \textbf{Formal Expression} & \textbf{Description} & \textbf{Evidence} \\
    \hline
    DSM-R1 & Moisture Bounds & \(0 \leq moisture \leq 100\) & DT is defined only for realistic moisture values & Static proof (state invariant) \\ 
    \hline
    DSM-R2 & Parameter Domain Validity & $\begin{aligned}
        & 0 < \text{soilType\_coeff} \leq  1 \hspace{0.2cm} \land \hspace{0.2cm} \text{surfaceArea\_m2} > 0 \hspace{0.2cm} \land \\
        & \text{soilCapacity} > 0 \hspace{1.15cm} \land \hspace{0.2cm} 0 \leq \text{infRate} \leq 1 \hspace{0.95cm} \land \\
        & 0 \leq \text{minT} \leq 100 \hspace{1.2cm} \land \hspace{0.2cm} 0 < \text{maxT} \leq 100 \hspace{0.8cm} \land \\
        & \text{minT} < \text{maxT} \hspace{1.45cm} \land \hspace{0.2cm} \text{pumpFlow\_L\_min} \geq 0 \end{aligned}$ & 
        DT only valid for meaningful physical parameters & Configuration validation \\ 
    \hline
    DSM-R3 & Environmental Stress Validity & \(evapFactor \geq 0\) & DT operates only under valid environmental stress & Input validation \\ 
    \hline
    DSM-V1 & Invariant Preservation & \(moisture(t) \in [0,100]\) & DT never generates invalid moisture & Static proof \\ 
    \hline
    DSM-V2 & Command Output Domain & \(optimizedCMD \in \{-1,0,1\}\) & DT output is always well-formed & Postcondition verification (static) \\ 
    \hline
    DSM-V3 & Control Logic Termination & OptimizeLogic() always terminates & DT control loop is total & Structural analysis \\ 
    \hline
    DSM-S1 & Decision-impact preservation & \(Inv_{Pot}(s) \land Execute(optimizedCMD) \Rightarrow Inv_{Pot}(s')\) & DSM can replace pot controller without violating safety envelopes & Static invariant reasoning under substitution \\
    \hline
    DSM-S2 & Feasibility Preservation & \(Feasible(state,optimizedCMD)\) & DSM never issues impossible commands or violates system constraints & State-dependent constraint analysis \\
    \hline
    DSM-F1 & Moisture Prediction Error & \(|moisture_{DT} - moisture_{PhysicalPot}| \leq \varepsilon\) & Quantifies closeness to physical pot & Empirical calibration + trace comparison \\ 
    \hline
    DSM-F2 & Evaporation Sensitivity Fidelity &
    \large \(|\frac{\partial moisture_{DT}}{\partial evapFactor} - \frac{\partial moisture_{Pot}}{\partial evapFactor}| \leq \gamma\) 
    & DT reacts similarly to environment within tolerance $\gamma$ & Empirical trace comparison \\ 
    \hline
    DSM-F3 & Decision Stability & $\begin{aligned}
    & |\Delta moisture| < \delta \Rightarrow \\
    & optimizedCMD(t) = optimizedCMD(t-1)
    \end{aligned}$ & DT avoids oscillatory decisions for small $\delta$ & Empirical trace comparison \\ 
    \hline
  \end{tabular}
    \label{DSMqualities}
\end{table}
\end{landscape}

\begin{landscape}
    \begin{table}[h]
    \centering
    \small
    \caption{Environment Digital Shadow (EDS) – Property Classification}
    \begin{tabular}{@{}|
    >{\raggedright\arraybackslash}m{1.2cm}|
    >{\raggedright\arraybackslash}m{2.8cm}|
    >{\raggedright\arraybackslash}m{7.2cm}|
    >{\raggedright\arraybackslash}m{3.8cm}|
    >{\raggedright\arraybackslash}m{3.6cm}|
@{}}
    \hline
    \textbf{No.} & \textbf{Property} & \textbf{Formal Expression} & \textbf{Description} & \textbf{Evidence} \\
    \hline
    EDS-R1 & Sensor Domain Coverage & $\begin{aligned}
        & temp \in [-10,50] \hspace{0.5cm} \land \\
        & hum \in [0,100] \hspace{0.8cm} \land \\
        & light \in [0,100] \end{aligned}$ & 
        Environment model valid only for realistic inputs & Input validation \& Static proof (state invariants) \\
    \hline
    EDS-V1 & Invariant Preservation & $\begin{aligned}
        & -10 \leq temp \leq 50 \hspace{0.5cm} \land \\
        & 0 \leq hum \leq 100 \hspace{1cm} \land \\
        & 0 \leq light \leq 100\end{aligned}$ & 
        State variables remain in bounds & Runtime assertion \& Static proof \\ 
        \hline
        EDS-V2 & Evaporation Factor Non-negativity & \(evapFactor \geq 0\) & Evaporation stress is physically meaningful & Postcondition verification (static) \\ 
        \hline
        EDS-V3 & Step Termination & Step() always terminates & Environment always progresses & Structural analysis \\ 
        \hline
        EDS-V4 & Environmental Consistency & \(same\ inputs \Rightarrow same\ evapFactor\) & Deterministic behaviour & Static determinism reasoning \\ 
        \hline
        EDS-F1 & Sensor fidelity bounds & $\begin{aligned}
        & |temperature_{physical} - temperature_{EDS}| \leq \varepsilon_1 \hspace{0.2cm} \land \\
        & |humidity_{physical} - humidity_{EDS}| \leq \varepsilon_2 \hspace{1.15cm} \land \\
        & |light_{physical} -light_{EDS}| \leq \varepsilon_3 \end{aligned}$ & 
        Quantifies closeness to real environment sensor readings & Empirical trace comparison \\ 
        \hline
        EDS-F2 & VPD Physical Fidelity & \(evapFactor \approx VPD \times light\) & EDS approximates physical evapotranspiration & Empirical calibration \\ 
        \hline
  \end{tabular}
    \label{EDSqualities}
\end{table}
\end{landscape}

\subsection*{Pot Model Instantiation}
Relevance is expressed through physical domain constraints. For example, in Table~\ref{Potqualities}, Property P-R1 (Moisture Bounds) is encoded directly as a state invariant which defines the admissible state space and may be expected to contribute to a domain-relevance argument.

\begin{vdmtt}
instance variables
  moisture: real := 50.0;
inv moisture >= 0.0 and moisture <= 100.0;
\end{vdmtt}

Invariant preservation and termination of operations contribute to verifiability. This can be achieved using formal proof obligations (POs), which ensure that any reachable state satisfies the invariant and guarantee the totality of operations by satisfaction of pre- and post-conditions. 

\begin{vdmtt}
private UpdateMoistureFromSensor: () ==> ()
UpdateMoistureFromSensor() == (
    -- update logic
) pre moistureSensor.getValue() >= 0.0 and
      moistureSensor.getValue() <= 100.0
post moisture >= 0.0 and moisture <= 100.0;
\end{vdmtt}

In this case study, there are no loops or recursion, hence termination verification is trivial. Nevertheless, we include these properties for completeness and to illustrate that even simple DTs can be subjected to systematic verification obligations.


Substitutability is interpreted as behavioural agreement with the physical pot under identical environmental inputs, including agreement in moisture trends (P-S1), consistency in threshold evaluation (P-S2), and agreement in decisions under identical inputs (P-S3). These may be validated via relational trace comparison. Fidelity properties concern the bounded deviation between simulated and measured moisture trajectories.

\subsection*{Decision Service Model Instantiation}
Properties relevant to the DSM are outlined in Table~\ref{DSMqualities}. Admissible domain constraints (e.g., valid moisture range and parameter domains) relate to Relevance. Verifiability includes well-formedness of output.

\begin{vdmtt}
private ComputeOptimisedCommand: () ==> ()
ComputeOptimisedCommand() == (
  -- Computation of Optimised Command
  -- ( -1 = nil, 0 = turn off, 1 = turn on )
) pre moisture >= 0.0 and moisture <= 100.0
post optimizedCommand.getValue() in set {-1,0,1};
\end{vdmtt}

The DSM differs architecturally from the Pot model; it is not intended to replicate controller behaviour but to optimise within safe bounds. Thus, Substitutability for the DSM may be defined in terms of decision-impact preservation (DSM-S1) rather than behavioural equivalence. In other words, if the DSM replaces the threshold controller, system safety invariants remain satisfied. Feasibility preservation (DSM-S2) additionally ensures that issued commands respect actuator and physical constraints. Fidelity properties (DSM-F1-F3) compare the predicted moisture evolution with the physical pot and ensure decision stability.

\subsection*{Environment DS Instantiation}
Properties relevant to the EDS are outlined in Table~\ref{EDSqualities}. The EDS abstracts environmental context. Relevance is defined through sensor-domain constraints (temperature, humidity, light ranges). Verifiability includes deterministic behaviour and non-negativity of derived evapotranspiration factors. Substitutability is not considered for the EDS because, by definition, a DS does not influence the physical system - it only provides context. Therefore, the question of substituting the DS for the real environment does not arise. 
Fidelity, however, remains necessary: the Pot DTs’ recommendations depend on accurate environmental data. Hence, Fidelity properties (EDS‑F1, EDS‑F2) quantify the deviation between simulated and measured environmental indicators.

\section{Implications for Compositional Verification}
\label{sec:compositionIssues}
In the case study to date, we have enumerated significant properties related to individual DT artefacts and indicated how they may play a role in verifying high-level qualities. We have not so far examined composition in an SoS setting. In this section, we share some insights from the case study so far on challenges that arise under composition. Local satisfaction of DT qualities does not guarantee their preservation at the SoS level.


First, fidelity becomes an aggregate property. Even if individual DTs satisfy local error bounds, their interactions under shared resource constraints (e.g., water allocation across multiple pots) may amplify deviations, leading to emergent inaccuracies at the SoS level. Second, relevance shifts in scope. At the constituent system level, relevance is defined in terms of physically meaningful state bounds. At the SoS level, relevance must additionally capture contribution to global objectives such as sustainability, efficiency, or resilience. Third, substitutability becomes relational and architectural rather than local. A DSM that is locally safe may still produce decisions incompatible with other DTs when composed. Thus, substitutability evolves from physical equivalence to behavioural compatibility across interacting services.

 
Performance and scalability concerns also arise with composing multiple DTs with real‑time synchronisation (e.g., each Pot DT reading the shared EDS, the SoS coordinator reading from all Pot DTs and the EDS then responding with priority-based commands). Communication latency, state‑synchronisation overhead, and potential co‑simulation bottlenecks are also introduced. While our FMU‑based export enables co‑simulation, the overhead of verifying properties across, say, 100 Pots instead of 3 would be nontrivial. Each DT’s verification obligations (invariant checks, trace comparisons) would multiply, and global property monitoring may require time-consuming techniques. Thus, scalability is not merely a runtime concern but a verification challenge too: compositional proof strategies must consider how verification cost grows with the number of DTs. 

These observations indicate that compositional verification must address both semantic (fidelity amplification, scope shift) and practical (performance, scalability) dimensions.
Local DT guarantees must serve as formal assumptions for higher-level reasoning, and additional verification obligations emerge at each layer of composition. This motivates future research into compositional proof strategies and contract-based reasoning across DT-enabled SoS architectures.

\section{Discussion}
\label{sec:discusssion}

This section interprets the results of our first examination of the potential for operationalising DT qualities, considering what the formal approach reveals about DT verification practices.

The operationalisation of DT qualities as structured sets of verifiable properties suggests patterns in how verification obligations depend on artefact roles in the SoS setting. Across all three artefacts, relevance properties consistently relate to state invariants verifiable through static proof, suggesting that domain appropriateness might be established early in the development lifecycle through type systems and invariant checking. This uniformity contrasts with substitutability and fidelity, where interpretation varies dramatically by architectural role.

For the Pot Model, substitutability manifests as behavioural agreement under identical inputs—the model must predict the same qualitative effects as the physical pot. For the DSM, substitutability instead means decision-impact preservation: the DSM need not replicate the physical controller's exact threshold logic behaviour, but must ensure that safety invariants remain satisfied when its recommendations are executed. This distinction is architecturally significant: it enables the DSM to provide optimised control strategies while the physical controller retains ultimate safety authority. This separation of concerns proves essential for modular verification in safety-critical DT applications.

The aggregation of properties across quality dimensions reveals dependencies that compositional verification must address. The Pot Model's fidelity-related properties (P-F1 to P-F4), see Table~\ref{Potqualities}, depend on the EDS's accuracy via the evapFactor input. This dependency is currently implicit in our verification approach; we assume the EDS provides bounded-error environmental data, but do not formally capture this assumption as a contract. Future compositional verification must make such dependencies explicit, enabling modular reasoning in which component verification relies on stated assumptions about its context.

The challenges identified in Section~\ref{sec:compositionIssues} suggest that verification of DT qualities at the SoS level requires more than simply aggregating local guarantees and errors. Instead, compositional fidelity reasoning must account for interaction patterns, shared resource allocation strategies, and timing dependencies.

Our integration of VDM-RT with FMI co-simulation bridges a gap between formal verification and industrial practice. Recent FMI-based work~\cite{2022fmi3} has focused on technical co-simulation improvements (clocked simulation, algebraic loop handling) but not on preserving formally verified quality properties through the FMU export process. Our demonstration that VDM-RT's formal constructs (invariants, contracts, termination guarantees) can be exported as FMUs while maintaining verification evidence provides a practical pathway for deploying formally verified DTs in industrial co-simulation workflows.

\section{Threats to Validity}
\label{sec:validity}

As with any case study, several threats to validity must be acknowledged.

\textbf{Internal validity.}
The VDM-RT models are validated through simulation and controlled execution traces rather than deployment on physical hardware. This enables observation and analysis but abstracts away sensor noise, actuator delays, and hardware imperfections that may affect real-world behaviour. The architectural separation between safety enforcement (physical controller) and DT-based optimisation (DSM recommendations) mitigates this threat by ensuring that DT-induced errors are bounded by physical safety constraints.

\textbf{Construct validity.}
The four DT qualities of relevance, verifiability, substitutability and fidelity are drawn from Oakes et al.~\cite{Oakes2023Qualities} based on recurring themes in DT literature. However, other quality dimensions (e.g., maintainability, explainability, security) may be equally important in specific application contexts. The property decompositions presented reflect our interpretation of these qualities for the greenhouse domain. Alternative decompositions may be appropriate for domains with different characteristics or stakeholder priorities.

\textbf{External validity.}
The greenhouse case study exhibits specific characteristics (shared environment, hierarchical control). Other domains (e.g., manufacturing, energy) may introduce additional complexities (e.g., safety‑critical real‑time constraints, legacy system integration) not captured here. Therefore, the generalisability of our quality‑instantiation patterns requires empirical validation through further case studies.


\section{Conclusion}
\label{sec:conclusion}

This paper has addressed the challenge of formally verifying DT qualities in SoS settings by considering the operationalising of abstract quality concepts -- relevance, verifiability, substitutability, and fidelity -- as sets of formally verifiable properties instantiated in VDM-RT models.

Our first primary observation is that meaningful quality decomposition is artefact-dependent: the same quality (e.g., substitutability) manifests as behavioural agreement for physical abstractions, decision-impact preservation for optimisation services, and safe context replacement for environmental models. This role-dependency has direct implications for modular verification strategies in multi-artefact DT systems. Our second observation is that VDM-RT's constructs potentially relate to verification evidence types (static proof, contract verification, trace comparison, empirical calibration).
Third, in the case study, we have observed the practical integration of formal verification with industry-standard co-simulation through FMI-compliant FMU export, bridging the gap between rigorous formal analysis and industrial simulation workflows.

Through the greenhouse case study, we have identified three specific challenges affecting DT composition: fidelity error amplification under resource constraints, the need for scope-shifting relevance definitions at the SoS level, and the evolution of substitutability from physical equivalence to architectural compatibility. These observations indicate that compositional verification of DT qualities requires explicit formalisation of dependencies between artefacts and the development of proof rules for deriving SoS-level quality guarantees from guarantees at the level of the constituent systems.

Our experience suggests that future work should address three key directions. \textit{SoS-level formalisation}: The full formalisation of the SoS-level DT, including multi-objective optimisation under global resource constraints, requires extending our quality framework to collective behaviours and emergent phenomena. \textit{Tool support and automation}: Increasing proof automation for invariant preservation, termination analysis, and trace comparison would significantly enhance scalability. Integration with existing VDM-RT toolchains such as Overture \cite{Larsen&10a} and VDM2C \cite{Hasanagic&19} should be prioritised. \textit{Additional case studies}: Empirical validation in domains such as smart manufacturing, energy systems, or transportation would strengthen generalisability claims and identify domain-specific patterns.

In conclusion, this work proposes a foundation for rigorous quality verification of individual DT artefacts as a step towards compositional verification in SoS settings. By transforming informal quality notions into structured property sets with explicit verification evidence, we enable systematic reasoning about DT trustworthiness. The challenges identified suggest a research agenda for scaling formal DT verification from individual components to SoS architectures, where verified local guarantees must be preserved under hierarchical composition. 

\noindent\textbf{\ackname} We are grateful to Prasad Talasila for support in the case study and Kenneth Lausdahl for support with VDM-RT and its FMU connection. Note: AI-powered tools, specifically ChatGPT and Grammarly, were used for clarity and text flow refinement in the production of this paper. The authors conducted a thorough review and necessary revisions and take full accountability for the final published material. 

\noindent\textbf{\discintname} The authors have no competing interests to declare.

\bibliographystyle{splncs04}
\bibliography{references}

\end{document}